\documentclass[a4paper,12pt]{article}
\usepackage{amsmath}
\usepackage{amsfonts}
\usepackage{amssymb}
\usepackage{marvosym}
\usepackage{wasysym}
\usepackage{caption}
\usepackage{subcaption}
\usepackage{fullpage}
\usepackage[]{graphicx}
\usepackage{xcolor}
\usepackage{hyperref}

\begin{document}

\begin{titlepage}
\title{Resonant high energy graviton to photon conversion at the post recombination epoch }
\author{Alexander D. Dolgov$^{1,2,3}$ and Damian Ejlli$^{1,4}$}
\date{}
\maketitle
\begin{center}
$^{1}$\emph{Dipartimento di Fisica e Scienze della Terra, Universit\'{a} degli Studi di Ferrara, 
Polo Scientifico e Tecnologico-Edificio C, Via Saragat 1, 44122 Ferrara, Italy}\\
$^{2}$\emph{Department of Physics,Novosibirsk State University, Pirogova 2, Novosibirsk 630090, Russia\\
$^{3}${ITEP}, Bol. Cheremushkinskaya 25, Moscow 117218 Russia} \\
$^{4}$\emph{Astroparticule et Cosmologie (APC),
Universit\'{e} de Paris Diderot-Paris 7\\
10, rue Alice Domon et L\'{e}onie Duquet,
75205 Paris Cedex 13
France}
\end{center}
\thispagestyle{empty}

\begin{abstract}
Resonant conversion of high energy gravitons into photons in large scale cosmological magnetic fields
at the post-recombination epoch is considered. It is shown that the probability of  the resonance photon 
production is much higher than the nonresonant one. As a result, an observable isotropic background 
of cosmic gamma rays might be created. As shown in our previous paper, an early population of primordial 
black holes  prior to big bang nucleosynthesis could be an efficient source of high frequency gravitational 
waves. For the primordial black hole mass about $10^8$ g the produced photons would be  the dominant 
component of the soft to hard cosmic x-ray background and for lower masses the spectrum is shifted 
down to the ultraviolet and optic.
\end{abstract}

\vspace{5cm}
\Email{Alexander D. Dolgov:  \href{mailto:dolgov@fe.infn.it}{\nolinkurl{dolgov@fe.infn.it}} }\\

\Email{Damian Ejlli: \href{mailto:ejlli@fe.infn.it}{\nolinkurl{ejlli@fe.infn.it}}}

\end{titlepage}

\pagenumbering{arabic}

\section{Introduction}\label{sec:1}

Observation of gravitational waves (GWs), which were generated in the early Universe, could bring invaluable information about physical processes at cosmological epochs from which no other messenger was able to come to us undisturbed. In a sense GWs are similar to the cosmic microwave background (CMB) photons which present a snapshot of the Universe at redshift $z\sim 10^3$. Evidently gravitational waves penetrate to the present day Universe from the much deeper past. Unfortunately it is extremely difficult to register them. At the present time there operate several high sensitivity detectors, such as ground based interferometers Virgo, LIGO, etc and some more are in progress (LISA, DECIGO, etc), dedicated to observation of very long gravitational waves with frequencies in the range of a fraction of a hertz up to 10 Hz. Such detectors might in principle register low frequency GWs from the inflationary epoch, as well as from catastrophic processes in the contemporary Universe, for a review see~\cite{gw-rev}. 

However, there exist physical processes which could efficiently generate very high frequencies of GWs far beyond the sensitivity range of the usual types of detectors. Cosmological mechanisms of production of such high frequency GWs are discussed in ~\cite{hi-f-gw}. In paper~\cite{ad-de-2} (hereafter we refer to it as paper 1) we considered a way to "see" very high frequency GWs through the process of the graviton to photon conversion in external magnetic fields. In the first paper~\cite{gertsen},  dedicated to this subject, the inverse process of transformation of electromagnetic waves into gravitational wave was studied, while later there appeared several more where the transition of GWs to electromagnetic radiation was considered~\cite{g-to-gamma}. In what follows,  we will use the results of~\cite{Raffelt:1987im} and of paper 1. In the last paper we have studied the transformation of GWs with high frequency (but still below electron mass) into electromagnetic radiation in cosmological magnetic fields after the hydrogen recombination epoch. We have shown that the transition probability is much larger, 
by more than by an order of magnitude, than the naive estimation, though the graviton energy was not high enough to reach 
an analogue of the Mikheyev-Smirnov-Wolfenstein resonance.  In this work we abandon this restriction and consider resonance graviton to photon transformation.

\section{Equation of motion of the graviton-photon system}

The total action describing the interaction between gravitational field and electromagnetic field is given by the sum of two terms: 
\begin{equation}\label{totalaction}
  S= S_g+S_{em}.
\end{equation}
The first term, $S_g$, is the Einstein-Hilbert action, given by 
\begin{equation}\label{gravaction}
S_g=\frac{1}{\kappa^2}\int d^4x\sqrt{-g} R,
\end{equation}
where $R$ is the Ricci scalar, $g=\mathrm{Det}(g_{\mu\nu})$ is the metric tensor determinant,
and $\kappa^2=16\pi G_N \equiv 16\pi / m_{Pl}^2$  (here and in what follows, the natural units, $c=\hbar=k_B=1$
are used). The amplitude of gravitational wave $h_{\mu\nu}$ is defined according to the equation:
\begin{equation}
\label{metricsplit}
 g_{\mu\nu}=\eta_{\mu\nu}+\kappa h_{\mu\nu}(\mathbf{x}, t),
\end{equation}
where $\eta_{\mu\nu}$ is the background flat space metric tensor and $h_{\mu\nu}$ are small perturbations around it, $|h_{\mu\nu}|\ll 1$.
In what follows we neglect curvature effects and only include an impact of the overall cosmological expansion on the evolution of
the mixed graviton-photon wave function or to be precise, on the density matrix.

The electromagnetic part of the action is equal to:
\begin{equation}
S_{em}=-\frac{1}{4}\int d^4x\sqrt{-g}\,F_{\mu\nu}F^{\mu\nu}
+ \frac{\alpha^2}{90m_e^4}\int\mathrm d^4x\sqrt{-g}\,
\left[(F_{\mu\nu}F^{\mu\nu})^2+\frac{7}{4}(\tilde{F}_{\mu\nu}F^{\mu\nu})^2 \right],
\label{em-action}
\end{equation}
where $\alpha =1/137$ and $m_e$ is the electron mass.
The first term here is the usual free  Maxwell action and  the second quartic one is the 
well-known Euler-Heisenberg contribution~\cite{EH}. This term describes the photon-photon elastic scattering 
or photon propagation in external electromagnetic field. Keeping in mind the latter possibility, we expand the total
electromagnetic  tensor as  
\begin{equation}
F_{\mu\nu}=F_{\mu\nu}^{(e)}+f_{\mu\nu},
\label{F-tot}
\end{equation}
where $F_{\mu\nu}^{(e)}$ is the external field tensor and $f_{\mu\nu}$ describes a propagating electromagnetic wave. 
The calculations below are done in the lowest essential order in the electromagnetic field strength and in the first order
in $h_{\mu\nu}$. The terms, which are quadratic in  $ F_{\mu\nu}^{(e)}$,  correspond to the Maxwell equation for the
external electromagnetic field. The terms quadratic in $f_{\mu\nu}$ describe propagating electromagnetic waves.  
The terms, quadratic in both $F_{\mu\nu}$ and $f_{\mu\nu}$ describe refraction index of photons in external electromagnetic
(in our case, magnetic) field. The graviton to photon transformation in the external magnetic field is described by the terms 
which are linear in $h_{\mu\nu}$, which comes from the metric, and in $f_{\mu\nu}$ and $F_{\mu\nu}$.

The Euler-Heisenberg effective Lagrangian is valid in the limit of low photon energy,
 when the kinematic variables $s= (k_1+ k_2)^2$ and $|t| = |(k_1-k_3)^2|$ are much smaller than $m_e^2$.
Here $k_1$ and $k_2$ are 4-momenta of the initial photons and $k_3$ and $k_4$ are the final ones. 
It  essentially means that the Euler-Heisenberg approximation is valid for sufficiently low photon energy, $\omega \ll m_e$ in the center of mass. For the photon scattering in the external (magnetic) field, which we consider in what follows, the restriction is much milder, $\omega$ may be much larger than $m_e$ in the laboratory frame.
The amplitude of the graviton-photon transformation depends upon the photon refraction index which in turn is proportional to the amplitude of the forward scattering of photons in the medium. We assume that the external magnetic field is homogeneous at macroscopically large scale $\lambda_B$ and slowly varying
with characteristic time $t_B$. It means that the energy and momentum transfer to or from
the external field by an order of magnitude are, respectively,
 $1/t_B$ and $1/\lambda_B$. 
Correspondingly the characteristic value of effective kinematical variable $s$ is of the order of
$ s \sim ( \omega/\lambda_B + \omega/t_B) $, so even for $\omega \gg m_e$ the Euler-Heisenberg effective Lagrangian would be valid. Thus for 
the study of the graviton-photon transformation we can apply equations used in~\cite{ad-de-2,Raffelt:1987im} in the Euler-Heisenberg approximation.

In the limit of high graviton frequency the wave equation of the mixed graviton-photon system  is accurately described by
the eikonal approximation and, hence, it is reduced to the first order matrix equation~\cite{Raffelt:1987im}:
\begin{equation}
\left[(\omega+i\partial_{\mathbf{x}})\mathbf I+
\begin{bmatrix}
  \omega(n-1)_{\lambda} & B_{T}/m_{Pl} \\
  B_{T}/m_{Pl}  & 0  \\
   \end{bmatrix}
\right] 
\begin{bmatrix}
  A_\lambda({\mathbf{x}}) \\
  h_\lambda({\mathbf{x}}) 
 \end{bmatrix}
=0\,,
\label{matrix}
\end{equation}
where $\mathbf I$ is the unit matrix, $\mathbf x$ is the direction of the graviton/photon propagation,
$n$ is the total refraction index of the medium, $\omega$ is the graviton energy, $B_T$ is the strength of the transverse external magnetic field $\mathbf B_e$ and $h_\lambda, A_\lambda$ are  respectively the graviton and photon wave functions  with $\lambda$ being the polarization index (helicity) of the graviton and the photon states. In the case of photons $\lambda=+$ indicates a polarization state perpendicular to the external magnetic field and $\lambda=\times$ indicates a state with polarization parallel to the external field. Note that graviton-photon transition is induced by the off-diagonal term in the effective potential matrix, which is equal to $B_T/m_{Pl}$, while the equation is diagonal with respect to different helicity states of graviton and photon (for their definition see~\cite{ad-de-2,Raffelt:1987im}).

The refraction index of gravitons is of course equal to unity, because of very weak interaction of gravitons with the medium where they  
propagate. The photon refraction index consists of three terms:
\begin{equation}
\label{totalindex}
n=n_\textrm{plasma}+n_\textrm{QED}+n_\textrm{CM},
\end{equation}
where the first term describes the interaction of photons with the electronic plasma; $n_\textrm{plasma}=-\omega_\textrm{pl}^2/2\omega^2$ is the refraction index due to plasma effects with the plasma frequency $\omega_\textrm{pl}^2=4\pi\alpha n_e/m_e$; where $n_e$ is the plasma density and should not be confused with the refraction index and $n_\textrm{CM}=n_\textrm{CM}^+-n_\textrm{CM}^\times=C\lambda_p B_e^2$ is the index of refraction due to 
the Cotton-Mouton effect that is  the double refraction of light in a liquid or gas in the presence of a constant transverse magnetic field, with $C$ being the 
Cotton-Mouton constant. 
The index of refraction associated with the vacuum polarization, $n_\textrm{QED}$, induced by 
a homogeneous magnetic fields has different expression in the case when the transverse magnetic field $B$ is greater or smaller than the critical magnetic field $B_c=m_e^2/e=4.41\cdot 10^{13}$ G. Moreover, it has two different values depending on 
the direction of the magnetic field with respect to the photon polarization vector. 
In the momentum space the modified Maxwell equations obtained from the total Lagrangian density of the electromagnetic field including one-loop QED effects, are
\begin{equation}
[k^2g_{\mu\nu}-k_\mu k_\nu+\Pi_{\mu\nu}(k, \omega, B)]A^\nu(k)=0.
\end{equation}
where $\Pi_{\mu\nu}$ is the vacuum polarization tensor. The general expression for the complex refraction index is given by \cite{Tsai:1974fa}:
\begin{equation}
\tilde n_{\times, +}(\omega)=\frac{k_{\times, +}}{\omega}=1-\frac{\Pi_{\times, +}}{2\omega^2},
\label{index}
\end{equation}
where $\Pi_{\times,+}$ are the eigenvalues of the photon polarization tensor. The usual index of refraction corresponding to dispersive phenomena\footnote{There is also a contribution to absorption indexes, $\kappa_{\times, +}$ 
due to production of $e^\pm$ pairs in magnetic fields for $\omega>4m_e$. However, for a large scale magnetic fields and 
in the case of $\omega\ll m_e(B_c/B)$, the absorption indexes are completely negligible since they decrease exponentially as 
$\kappa_{\times, +}\propto \exp{(-\omega B_c/m_eB)}$.} in a medium is given by the real part of 
equation \eqref{index}, namely $n=\textrm{Re}\{\tilde n\}$ and in the case of $\omega\ll (2m_e/3)(B_c/B)$ it reads \cite{Tsai:1974fa}
\begin{equation}\label{QED-index}
n_{\times, +} =1+\frac{\alpha}{4\pi}\left(\frac{B_T}{B_c}\right)^2\left[\left(\frac{14}{45}\right)_{\times}, \left(\frac{8}{45}\right)_{+}\right].
\end{equation}
where $B_T=B_e\sin\Theta$. We can see that equation \eqref{QED-index} is  
valid for a wide range of graviton energies and depends essentially on the strength of the background magnetic field. In fact, as we see in what follows, with the existing upper limit on the strength of the
magnetic field equal to a few $\cdot 10^{-9}$ G, 
equation \eqref{QED-index} is valid for the present day graviton energies $\omega\ll m_\textrm{Pl}$. 
Hence equation \eqref{QED-index} is valid for a wide range of graviton energies 
as far as we deal with large scale cosmological magnetic fields.

In this work the density matrix formalism is used because the oscillation length $l_{osc}$ is greater than 
or comparable to the mean free path $l_{free}$ ($l_{free} \lesssim l_{osc}$). The density matrix operator satisfies the 
Liouville-von Neumann equation
\begin{equation}
i\frac{d \hat\rho}{d t }= \hat H\hat\rho-\hat\rho\hat H^\dagger,
\label{densityoperator}
\end{equation}
where $\hat H$ is the total Hamiltonian operator of the system which is, in general, not Hermitian because the system is not closed 
due to interaction with the medium. In general the total Hamiltonian  is given by the sum of a Hamiltonian operator, $ \hat M$ and an 
anti-Hermitian term 
which describes an interaction of the system with the medium, $i \hat \Gamma$. The last term describes damping of  oscillations
due to scattering and absorption in the medium. 

\section{Mixing in  expanding universe}

The contribution of different physical processes into the photon damping strongly depends upon the photon
energy and the cosmological redshift, $z$. Before the hydrogen recombination i.e. at  $z>1090$~\cite{Komatsu:2010fb},
 the matter is 
almost completely ionized, while at smaller redshifts the ionization fraction asymptotically drops down almost 
to $10^{-5}$ till the period of the later reionization. At low photon energies, from a few eV up to a few keV, and after the
recombination, the most important role is played by the photoelectric effect, i.e.
the photon absorption with subsequent ionization of the atom.  We are interested in photons with energy much higher
than the atomic binding energy, $\omega \gg I = \alpha^2 m_e/2 = 13.6$ eV. For relatively low energy photons,
$\omega < m_e$, the cross section of this process is \cite{Landau:1978}
\begin{equation}
\sigma = \frac{2^8 \pi}{3} \alpha a^2 \left(\frac{I}{\omega}\right)^{7/2} ,
\label{sigma-ion1}
\end{equation}
where $a = 1/(m_e \alpha) \approx 0.53 \cdot 10^{-8} $ cm is the Bohr radius. For larger photon energy, i.e. for $\omega > m_e$ the cross section is
\begin{equation}
\sigma = \frac{2\pi \alpha^6}{m_e \omega}
\label{sigma-ion2}
\end{equation}
At the intermediate region,  $\omega \approx m_e$, both expressions are  quite close to each other
numerically.

As we have already mentioned, 
the photoelectric effect is the most important one in the energy range 
from a few eV up to a few keV. In the energy range from a  few keV up to a few MeV the Compton scattering is the dominant process 
which breaks the coherence of the graviton-photon oscillations. 
For higher energies it turns that pair production on atomic nuclei is the dominant process starting 
from few MeV up to 100 GeV. However, as we  see in the next section our consideration is limited to the
present day energy range from keV up to 100 keV. 
At the recombination time which is our starting point, 
the interval 1-100 keV corresponds to the energy range, roughly, 1-100 MeV. 
Evidently in this case the photoelectric effect is completely irrelevant
\cite{Heitler1954}.
The absorption of $\gamma$ rays by heavy elements at later stage due to photoelectric effect might be important but the cosmological density of heavy elements is low, so for this reason we neglect it~\cite{Cruddace1974} .

The total cross section due to the Compton scattering is given by the Klein-Nishina formula \cite{KN} and is equal to
\begin{equation}
\sigma_{KN}=\frac{3}{4}\sigma_{T}\left[\frac{2+x(1+x)(8+x)}{x^2(1+2x)^2}+\frac{(x^2-2x-2)\log(1+2x)}{2x^3}\right],
\label{totKN}
\end{equation}
where $x$ is the ratio of the photon energy to the electron mass, $x=\omega/m_e$ and $\sigma_T=6.65\cdot 10^{-25}$ cm$^2$ is the Thompson cross section. The total cross section of the pair production is not an easy task to find for almost all elements, however, for photon energies in the range $1\ll x\ll 1/\alpha Z^{1/3}$ there is an approximate expression which reads \cite{Heitler1954}
\begin{equation}\label{Cross-pp}
\sigma_{pp}=\frac{\alpha Z(Z+1)}{\pi}\sigma_T\left[\frac{28}{24}\ln(2x)-\frac{218}{72}\right].
\end{equation}
Equation \eqref{Cross-pp} takes into account the cross section of pair production in the nuclei field and in the electron field. Contribution from only atomic nuclei is proportional to $Z^2$ while if we take into account also pair production in the electron field one has to replace $Z^2\rightarrow Z(Z+1)$. 

As we have mentioned above, the damping term, 
which enters the Liouville-von Neumann equation, is diagonal in the helicity space and has a single entry in the graviton-photon matrix,
at the $\gamma\gamma$ matrix element: $\Gamma=\textrm{diag} [\Gamma_\gamma, 0]$, where
the damping coefficient is given by
\begin{equation}
\Gamma_\gamma=n_e\sigma_{KN}+\sum_in_i\sigma_{pp}^i .
\end{equation}
Here $n_e$ is the free electron number density, $n_i$ is the number density of the $i$-th atomic species, 
$\sigma_{KN}$ is the total cross section of the Compton scattering and $\sigma_{pp}$ is the total absorption cross section 
of the pair production. 

In order to take into account the Universe expansion we write the time derivative in von-Neumann equations as $\partial_t=Ha\partial_a$ where $H$ is the Hubble parameter and $a$ is the scale factor. With the mixing matrix $M$ presented below and that given by Eq. \eqref{matrix}, we can write
\begin{equation}
M=
\begin{bmatrix}
m_\lambda & m_{g\gamma}\\
m_{g\gamma} & 0
\end{bmatrix}
\end{equation}
where  $m_\lambda=\omega(n-1)_\lambda$ and $m_{g\gamma}=B_T/m_{Pl}$. The terms proportional to the matrix $\omega\mathbf{I}$ in Eq. \eqref{matrix} and a minus sign which arises when we write Eq. \eqref{matrix} into a Schr\"{o}dinger-like equation do not give contribution to the von-Neumann equations because they give an overall phase which can be absorbed in $\Psi_\lambda=[A_\lambda, h_\lambda]^\textrm{T}$ with T denoting a matrix transpose.
Correspondingly the evolution of the matrix elements is determined by the equations:
\begin{eqnarray}\label{densitysys}
\rho_{\gamma\gamma}' &=&\frac{-2m_{g\gamma}I - \Gamma_\gamma\, \rho_{\gamma\gamma}}{Ha},\\
\rho_{gg}' &=& \frac{2m_{g\gamma}I}{Ha} ,\\
R'&=& \frac{mI-\Gamma_\gamma R/2}{Ha}\label{R1} ,\\
I'&=& \frac{-mR-\Gamma_\gamma I /2 - m_{g\gamma}(\rho_{gg}-\rho_{\gamma\gamma})}{Ha},
\end{eqnarray}
where we split the off-diagonal terms of the density matrix as $\rho_{g\gamma}^{*}=\rho_{g\gamma}=R+iI$ 
with $R$ and $I$ being the real part  and the imaginary part respectively. 
The terms $m$ and $m_{g\gamma}$ which enter matrix $M$
can be written as functions of the
scale factor as follows (in units of cm$^{-1}$):
\begin{eqnarray}\label{system}
m_{\gamma g}(a) &=& 8 \cdot 10^{-26}\left[\frac{B_i}{1\textrm{G}}\right]\left[\frac{a_i}{a}\right]^2,\\
m_\lambda(a)&=&1.33 \cdot 10^{-27}\left(\frac{B_i}{1 \textrm{G}}\right)^2\left(\frac{\omega_i}
{1\textrm{eV}}\right)\left(\frac{a_i}{a}\right)^5-1.12 \cdot 10^{-14}X_e(a)\left(\frac{1\textrm{eV}}{\omega_i}\right)\left(\frac{a_i}{a}\right)^2,\nonumber
\end{eqnarray}
where $B_i$ is the value of the magnetic field at recombination time $t_i$, $a_i$ is the scale factor at recombination time, $\omega_i$ is the initial graviton energy at recombination and $X_e(a)$ is the ionization fraction of free electrons at the post-recombination epoch. Here we have neglected the contribution of Cotton-Mouton effect to 
the refraction index due to difficulties in determining the Cotton-Mouton constant $C$ at the post-recombination plasma. 
Deriving eqs. \eqref{system} we have expressed the free electron density as $n_e=X_en_B$ where $n_B$ is the total baryon 
number density. The baryon number density can be written as a function of temperature $T$ as $n_B=n_B(t_0)(T/T_0)^3$ where $T_0=2.275$ K is the present day temperature of the CMB photons and the total baryon number density at present is $n_B(t_0)\simeq 2.47\cdot 10^{-7}$ cm$^{-3}$ \cite{Komatsu:2010fb}. The ionization fraction is not easy to calculate analytically and often 
numerical calculations are used for its determination. 
The evolution of  $X_e$ is determined by the following differential equation \cite{Weinberg:2008}:
\begin{equation}\label{eqioniz}
\frac{\mathrm d X_e}{\mathrm d a}=-\frac{\alpha n_B}{Ha}\left(1+\frac{\beta}{\Gamma_{2s}+8\pi/\lambda_{\alpha}^3n_B(1-X_e)}\right)^{-1}
\left(\frac{SX_e^2+X_e-1}{S}\right),
\end{equation} where $\Gamma_{2s}=8.22458$ s$^{-1}$ is the two-photon decay rate of $2s$ hydrogen state, 
$\lambda_{\alpha}=1215.682\cdot 10^{-8}$ cm is the wavelength of the Lyman $\alpha$ photons, $\alpha(T)$ is the case B recombination 
coefficient,
and $S(T)$ is the coefficient in the Saha equation, $X(1+SX)=1$. Coefficient $\alpha$ depends on the scale factor as \cite{Hummer1994}:
\begin{equation}
\alpha(a)=\frac{1.038\cdot 10^{-12}a^{0.6166}}{1+0.352a^{-0.53}},
\end{equation}
while  $S(a)$ is equal to
\begin{equation}
S(a)=6.221\cdot 10^{-19}e^{53.158a}a^{-3/2}.
\end{equation}
Coefficient $\beta$, which is also a function of temperature, can be expressed through $\alpha$ as follows
\begin{equation}
\beta(a)=3.9\cdot 10^{20}a^{-3/2}e^{-13.289a}\alpha.
\end{equation}
With these parameters Eq. \eqref{eqioniz} is solved numerically and the solution determines the number density of free electrons at the post-recombination epoch.

\begin{figure}[htbp]
\begin{center}
\includegraphics[scale=1.2]{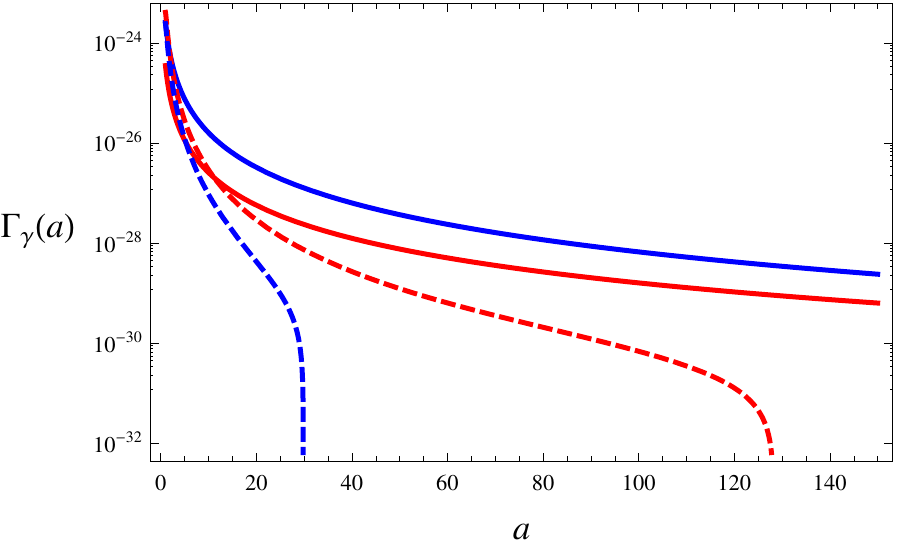}
\caption{$\Gamma_\gamma^{C}$ and $\Gamma_\gamma^{pp}$ as functions of  scale factor $a$. 
$\Gamma_\gamma^{C}$ (in red color) and $\Gamma_\gamma^{pp}$ (in red dashed color) are presented for
the initial graviton energy $\omega_i=10^9$ eV;  $\Gamma_\gamma^{C}$ (in blue color) and $\Gamma_\gamma^{pp}$ (in blue dashed color)
are presented  for the initial graviton energy $\omega_i=10^8$~eV.}
\label{fig:Fig1}
\end{center}
\end{figure}

The damping term due to the Compton scattering is given by (in units of cm$^{-1}$):
\begin{equation}
\Gamma_\gamma^C(a)= 1.6 \cdot 10^{-22}X_e(a)F(a; \omega_i)\left[\frac{a_i}{a}\right]^3,
\end{equation}
where $F(a; \omega_i)=F(x)$ with $F(x)$ being the expression within the square brackets in eq.~\eqref{totKN}. 
For high energy photons, when the photon wave length is smaller than the atom size, the photons equally well interact with free electrons and electrons bound in atoms. So for such an energy range the ionization fraction should be taken equal to unity, $X_e = 1$. However, in the refraction index due to plasma effects the ionization fraction is given from the solution of Eq. \eqref{eqioniz} with $X_e\ll 1$ because plasma effects are sensitive to only the number density of the free electrons at the post recombination epoch.
The total damping term which arises due to pair production is the sum of damping on hydrogen and helium nuclei. Noting that for hydrogen the atomic number is $Z=1$ and for helium $Z=2$ we can easily see that 
$\sigma_{pp}^{(He)}=3\sigma_{pp}^{(H)}$. The total number density per co-moving volume of hydrogen nuclei is $n_H\simeq n_B$ and the total number density of primordial atomic helium is $n_{He}\simeq Y_pn_B/4(1-Y_p)$ where $Y_p=4n_{He}/(4n_{He}+n_H)$ is the fractional abundance by weight of primordial helium. Taking this into account, 
we find that the damping term due to pair 
production on hydrogen nuclei and helium atoms is given by 
(in units of cm$^{-1}$):
\begin{equation}
\Gamma_\gamma^{pp}\simeq 9.85\cdot 10^{-25}\left(1+\frac{3}{4}\frac{Y_p}{Y_p-1}\right)G(a;\omega_i)\left[\frac{a_i}{a}\right]^3,
\end{equation}
where $G(a; \omega_i)=G(x)$ is the function within the 
square brackets in Eq. \eqref{Cross-pp}. In Fig.~\ref{fig:Fig1} the damping coefficients due to the Compton scattering and pair production,
$\Gamma_\gamma^{C}$ and $\Gamma_\gamma^{pp}$  respectively, are presented.
We can see that for the initial graviton energy 
$\omega_i=10^9$~eV, $\Gamma_\gamma^{pp}$ (red dashed color) is larger than $\Gamma_\gamma^{C}$ (red color) for $a<10$.
Here we use the 
normalization $a_i=1$ at the 
recombination time and $Y_p\simeq 0.24$. For $a>10$, $\Gamma_\gamma^{C}$ takes over $\Gamma_\gamma^{pp}$ which 
goes to zero at the beginning of the reionization epoch. In the case of graviton initial energy 
$\omega_i=10^8$~eV,  $\Gamma_\gamma^{pp}\sim \Gamma_\gamma^{C}$ for $a<5$, while for $a>5$, 
$\Gamma_\gamma^{pp}$ rapidly decreases to 
zero when the photon energy is below the threshold of 
pair production $\omega(a)<2m_e$. Therefore, $\Gamma_\gamma^{pp}$ due to pair production is important only for photon energies above $10^8$ eV 
and completely negligible for lower  energies.

\section{Photon production at post recombination epoch} \label{sec:3}

In the previous section we derived equations of motion for the elements of the density matrix $\rho$ in the case of  an 
expanding universe. In this section we 
solve them numerically and focus  on the post-recombination epoch. 
The key ingredients needed in order to solve Eqs. \eqref{system} are the comoving Hubble radius $Ha$, the strength of the 
magnetic field $B$, and the ionization fraction $X_e(a)$. At the post-recombination epoch, the Universe was 
dominated by non relativistic matter, while at the present time it is dominated by the vacuum energy. Hence we can write
\begin{equation}
Ha=H(t_\textrm{rec})[\Omega_M/a+\Omega_\Lambda a^2]^{1/2},
\end{equation}
where $\Omega_M\simeq 0.3$ is the present day matter density parameter, $H^{-1}(t_\textrm{rec})=6.7\cdot 10^{27}$ cm is the value of the Hubble 
distance at the recombination time, and $\Omega_\Lambda\simeq 0.7$ is the density parameter of the 
vacuum energy.  
In the case of redshift with respect to the present epoch, $z>1$ the contribution of the cosmological density parameter is negligible, because it becomes important only 
for $z\lesssim 1$; therefore, for $z>1$ we can take, $\Omega_M\simeq 1$ and for $z\lesssim 1$ we take $\Omega_M\simeq 0.3$ and $\Omega_\Lambda
\simeq 0.7$.

\begin{figure}[htbp]
\begin{center}
\includegraphics[scale=1.3]{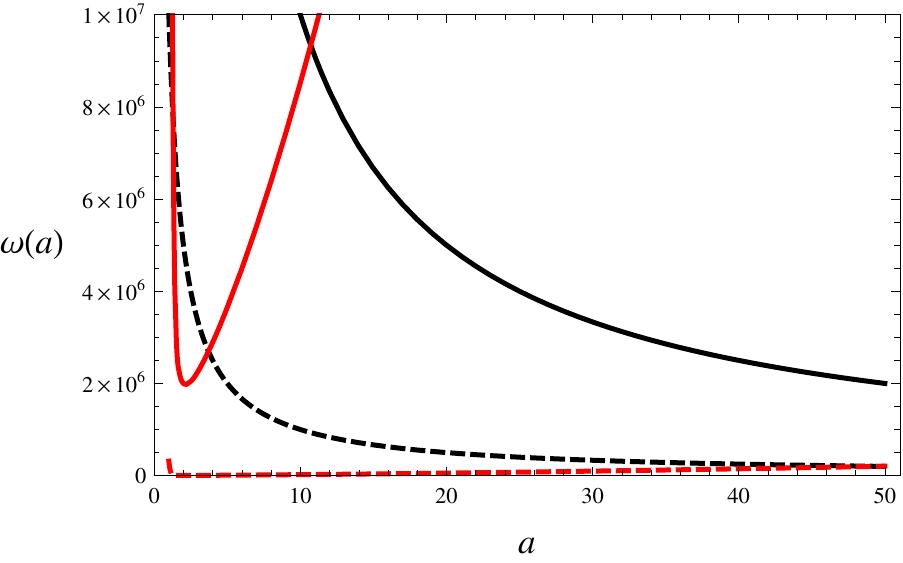}
\caption{Plot of the resonance frequency $\omega_\textrm{res}$ and graviton frequency $\omega$ as functions 
of scale factor $a$. In black color and in black dashed color 
the graviton frequencies are presented as  functions of $a$ for the initial graviton energies 
$\omega_i=10^8$ eV and $\omega_i=10^7$ eV respectively.
In red color and in red dashed color
the resonance frequencies are presented  for the initial values of the magnetic field $B_i=3\cdot 10^{-3}$ G and $B_i=1.2$ G
respectively.}
\label{fig:Fig2}
\end{center}
\end{figure}

For some particular value of the graviton energy in the range $\omega>m_e$ 
the oscillation of gravitons into photons is in resonance and the oscillation probability is strongly enhanced
The resonance occurred at the frequencies when $m_\lambda$, [Eq. (\ref{system})] vanishes.
At the resonance the mixing between the graviton and the photon reaches the maximum value of $\pi/4$.
The resonance energy as a function of the scale factor is given by
\begin{equation}\label{resonance-freq}
\omega_\textrm{res}(a)=2.9\, X_e^{1/2}(a)\left(\frac{1 \textrm{G}}{B_i}\right) a^{3/2}\; \textrm{MeV}.
\end{equation}
Given a graviton with the energy which drops down as a function of the scale factor  as $\omega(a)=\omega_i/a$, 
its energy would eventually cross the resonance energy when $\omega_\textrm{res}(a)=\omega(a)$. Since the resonance energy depends on the 
ionization fraction and due to the fact that an analytical form of the 
latter is in general unknown, we present
in Fig. \ref{fig:Fig2}  a graphical solution of the equation $\omega_\textrm{res}(a)=\omega(a)$. 
We can clearly see that the resonance is generally reached relatively early 
with respect to the recombination epoch and in some cases it is crossed twice as is the case of gravitons with the 
initial energy $\omega_i=10^7$ eV and the initial value of the magnetic field $B_i=3\cdot 10^{-3}$ G. 
In this paper we use the upper bounds on the present large scale magnetic fields obtained from the angular fluctuations of
CMB constraints, $B(t_0)\lesssim 3\cdot10^{-9}$~G~\cite{Paoletti:2012bb}
and from the measurements on the Faraday rotation of the 
CMB, $B(t_0)\lesssim 6\cdot 10^{-8}-2\cdot 10^{-6}$~G~\cite{Kahniashvili:2008hx}.

\begin{figure}[htbp]
\begin{subfigure}[b]{0.3\textwidth}
\centering
\includegraphics[scale=0.85]{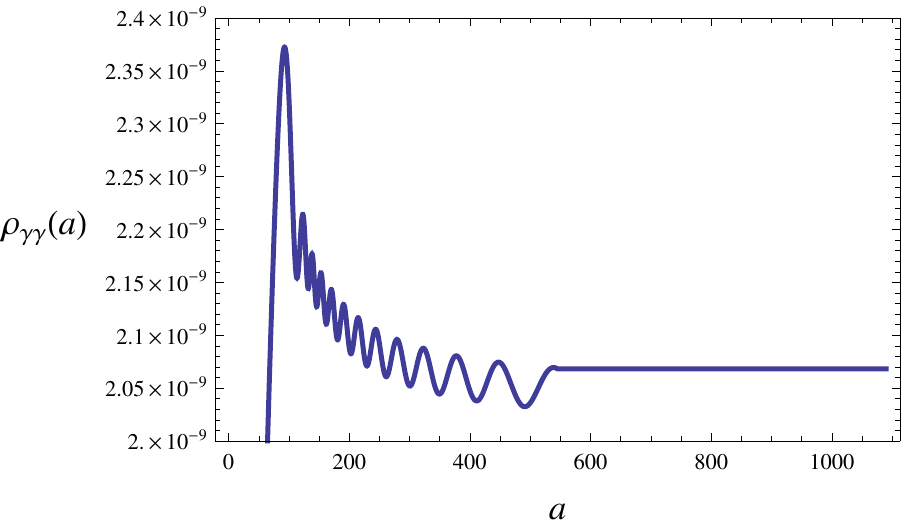}
\caption{}
\label{fig:Fig3}
\end{subfigure}\hspace{3.1cm}
\begin{subfigure}[b]{0.3\textwidth}
\centering
\includegraphics[scale=0.85]{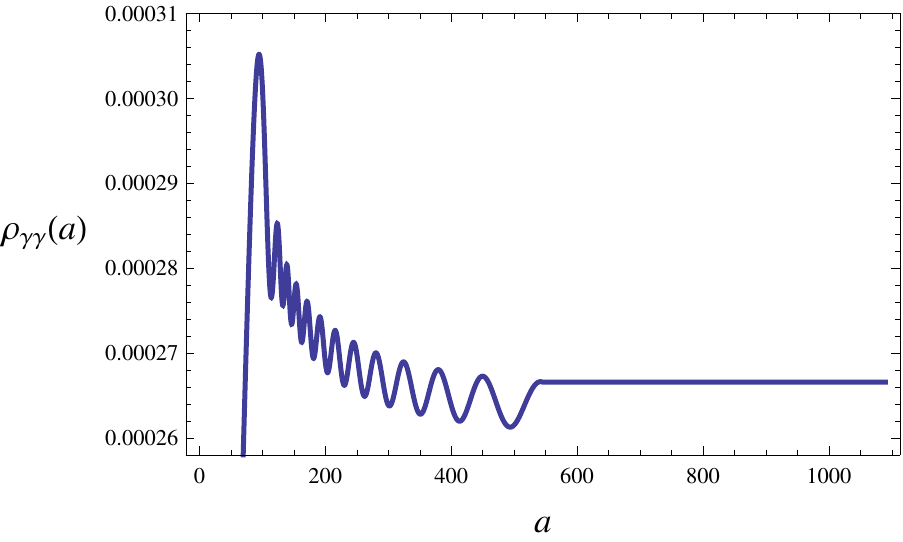}
\caption{}
\label{fig:Fig4}
\end{subfigure}     
\caption{The photon production probability as a function of the scale factor for initial graviton energy $\omega_i=10^7$ eV and the initial value of the magnetic field $B_i=3\cdot 10^{-3}$ G~(a) and $B_i=1.2$ G (b).}
\label{fig:Fig3a}
\end{figure}

The background of GWs at the
recombination epoch is assumed to be unpolarized and isotropic and, therefore,  the following
initial conditions $\rho_{\gamma\gamma}(a_i)=0$, $\rho_{gg}(a_i)=1/2$, $R(a_i)=0$, and $I(a_i)=0$ are imposed on 
solutions of Eqs. \eqref{system}. 
The factor $1/2$ in the initial conditions for the graviton probability takes into account the statistical weight of the polarization 
state $\lambda$. However, since we work with the reduced density matrix, namely we solve the evolution equation 
for a given polarization state $\lambda$, at the end we should multiply $\rho_{\gamma\gamma}$ by  the factor 2 in order to take into account that the initial background of GWs is composed of two independent polarization states. In Fig. \ref{fig:Fig3a}  the photon survival probability 
as a function of the scale factor $a$ starting from recombination epoch until the present epoch is presented. 
In the derivation, we took into account the Universe reionization, which according to \cite{Dunkley:2008ie} started at 
redshift $z\sim 20$ and finished with complete ionization at redshift $z\sim 7$. 
In Fig.~\ref{fig:Fig3a} it is clearly seen that after the onset of reionization at $a\sim 52$ the photon survival probability 
starts to oscillate and to decrease until the Universe becomes dominated by the vacuum energy at $a\sim 545$, 
and afterwards the probability remains practically constant.

The photon production probability depends upon
several factors such as the energy of initial gravitons and the matter content at the post- recombination epoch. 
As one can see in Fig. \ref{fig:Fig3a}(a), the photon production probability rapidly increases starting at $a \simeq 10$ for the initial value of magnetic field $B_i=3\cdot 10^{-3}$ G. In this case 
the resonance is crossed twice for $\omega(a)\simeq 7\cdot 10^6$ eV and  $\omega(a)\simeq 3\cdot 10^6$ eV see Fig. \ref{fig:Fig2}. For larger value of the  magnetic field, $B_i=1.2$ G, the increase starts at 
$a\simeq 50$, see  Fig. \ref{fig:Fig3a}(b) and the value of the graviton energy at the resonance cross is $\omega(a)\simeq 10^6$ eV. Such rise is partly explained by a rapid decrease of the ionization fraction which starts to grow at  $a \simeq 1$ and approaches asymptotically a constant value at $a \simeq 20$. 
Another factor leading to an increase of the production probability is the resonance effect due to a decrease of the 
graviton energy. 

\begin{figure}[htbp]
\begin{subfigure}[b]{0.3\textwidth}
\centering
\includegraphics[scale=0.85]{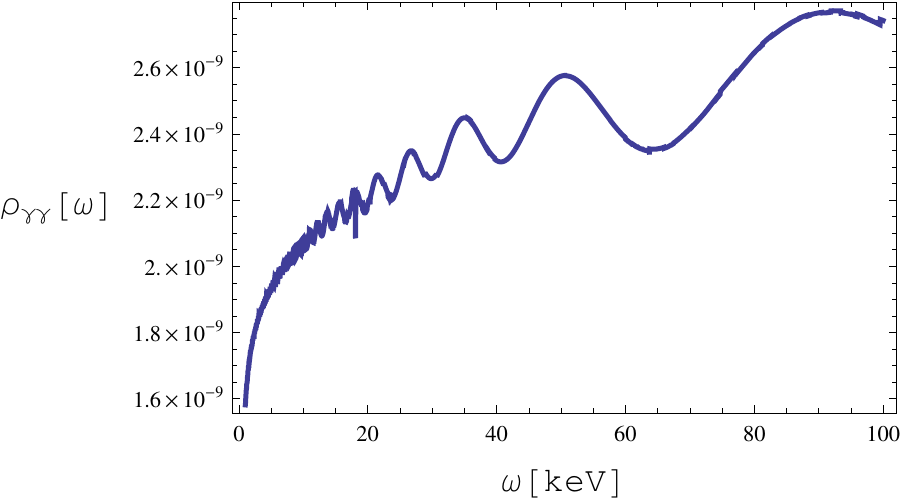}
\caption{}
\label{fig:Fig5}
\end{subfigure}\hspace{3.1cm}
\begin{subfigure}[b]{0.3\textwidth}
\centering
\includegraphics[scale=0.85]{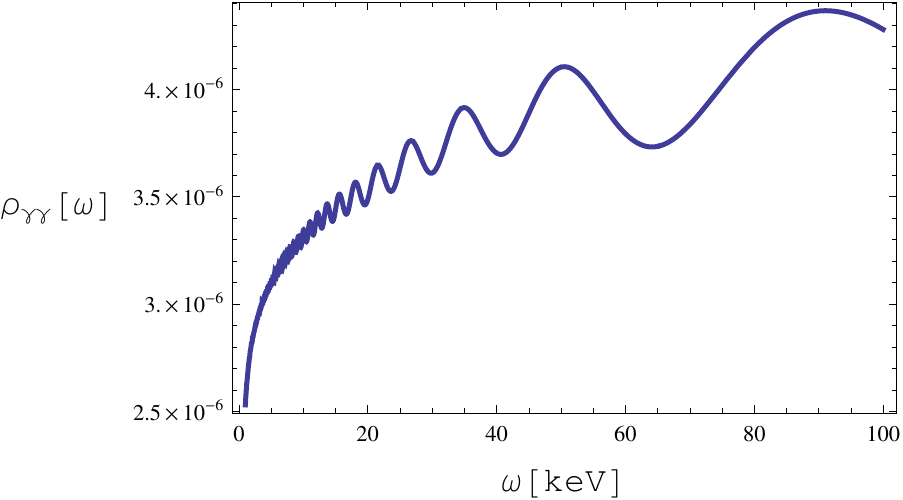}
\caption{}
\label{fig:Fig6}
\end{subfigure}     
\caption{Probability, $\rho_{\gamma\gamma}$, of the photon production by gravitons as a function of the
photon energy at the present cosmological epoch, $a=1090$, for the initial values of the magnetic field 
$B_i\simeq 3\cdot 10^{-3}$ G (a) and $B_i\simeq 0.12$ G (b).}
\label{fig:Fig4a}
\end{figure}

\begin{figure}[htbp]
\begin{subfigure}[b]{0.3\textwidth}
\centering
\includegraphics[scale=0.85]{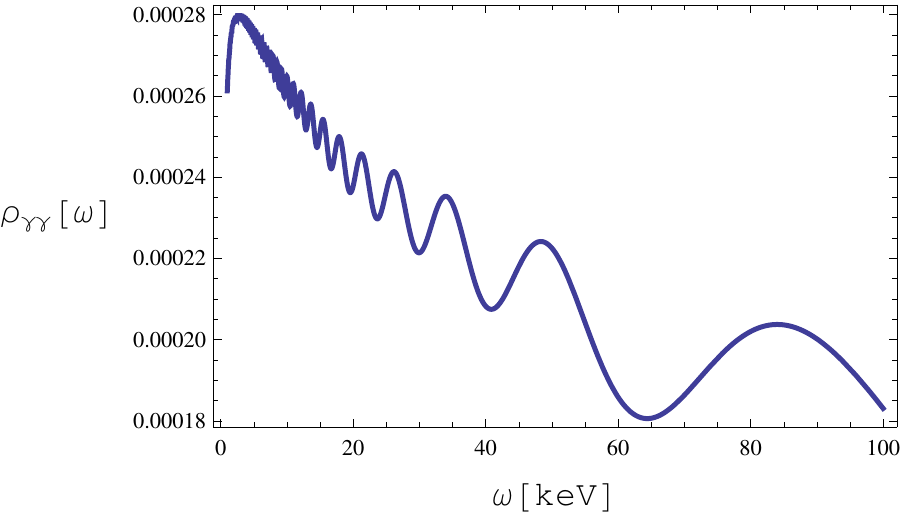}
\caption{}
\label{fig:Fig7}
\end{subfigure}\hspace{3.1cm}
\begin{subfigure}[b]{0.3\textwidth}
\centering
\includegraphics[scale=0.85]{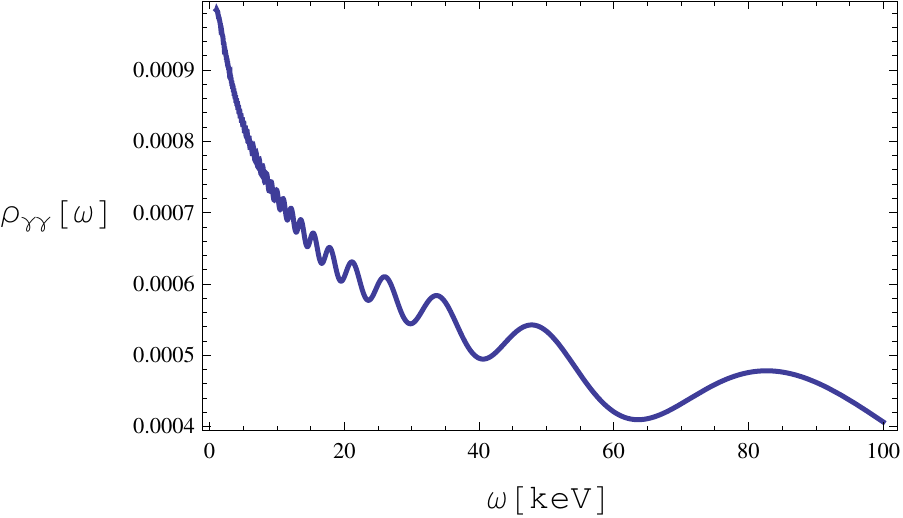}
\caption{}
\label{fig:Fig8}
\end{subfigure}     
\caption{Probability, $\rho_{\gamma\gamma}$, of the photon production by graviton as a function of the photon energy 
at the present cosmological epoch, $a=1090$, for the initial values of the magnetic field $B_i\simeq 1.2$ G (a) 
and $B_i\simeq 2.37$ G (b) .}
\label{fig:Fig5a}
\end{figure}

In Figs. \ref{fig:Fig4a} and  \ref{fig:Fig5a} the photon production probability is depicted as a function of the graviton or photon energy at present epoch, $a=1090$. 
The probability is much higher than the non-resonant one, roughly by 3 orders of magnitude.
The maximum value of the photon production probability strongly depends upon the initial value of the magnetic field 
taken at recombination. For $B_i>1$ G the probability reaches maximum value in
the frequency range $1-10$ keV and for $B_1<1$ G it is reached approximately at  $\omega \simeq100$ keV.

\section{Graviton production in the early universe. \label{s-GW-production}}

In paper 1 we considered several mechanisms of GW production in the early Universe and calculated the 
probability of the graviton-photon transformation in the high energy part of the graviton spectrum. According to
the model of \cite{hi-f-gw}, the density of the  GWs produced by primordial black holes (PBHs) 
is substantial and is concentrated in the energy range from eV up to keV. 
In paper 1 we estimated the amount of the produced photons for different values of large scale magnetic fields  
in energy part of the spectrum of soft x rays.
Here we present the calculations of the spectrum of the graviton-created photons for higher energies, 
namely in the energy range of several keV. 

As is well known, a black hole emits thermal spectrum of particles with masses $m \lesssim T_\textrm{BH}$. The 
luminosity with respect to this process is approximately (neglecting gray body corrections) equal to 
\begin{equation}
\frac{dE}{dtd\omega}=\frac{2N_{eff}}{\pi}\frac{M^2}{m_\textrm{Pl}^4}\frac{\omega^3}{\exp(\omega/T_\textrm{BH})\pm1},
\end{equation}
where $T_\textrm{BH}$ is the BH temperature, $N_{eff}$ is the effective number of species of the
emitted light  particles, and $\omega$ is the energy of the emitted particle.
Taking into account that the produced gravitons carry about 1\% of the total energy, we find that
the flux of GWs emitted by a BH with mass $M_\textrm{BH}$ at the present time is \cite{hi-f-gw}
\begin{equation}
F_\textrm{gw}(\omega; t_0)\simeq 8.35\cdot 10^{11}\left(\frac{\omega}{1 \textrm{keV}}\right)^4\left(\frac{10^5 \textrm{gr}}{M}\right)^2\left(\frac{N_{eff}}{100}\right)^2I\left(\frac{\omega}{T_\textrm{BH0}}\right)\, \left[\frac{\textrm{keV}}{\textrm{cm}^2\;\textrm{s}}\right].
\end{equation}
Here $T_\textrm{BH0}$ is the  temperature of an evaporating BH, rescaled to the  present time,
and we use the fact that for a stochastic background of GWs the energy flux is given by 
$F_\textrm{gw}=cT_{00}^\textrm{gw}=c\rho_c\Omega_\textrm{gw}$ where $T_{00}^\textrm{gw}$ is the energy density
of GWs and $\rho_c\simeq 5.27$ keV/cm$^3$ is the critical energy density. 
The function $I(\omega/T_\textrm{BH0})$ is an integral over redshift from the time of PBH formation, until its complete 
evaporation and is equal to
\begin{equation}\label{integral}
I\left(\frac{\omega}{T_\textrm{BH0}}\right)=\int_{0}^{z_\textrm{max}} \frac{dz(1+z)^{1/2}}{\exp[(1+z)\omega/T_\textrm{BH0}]-1},
\end{equation}
with $1+z_{max}$ given by
\begin{equation}
1+z_{max}=3.58\cdot 10^{14}\left(\frac{100}{N_{eff}}\right)^{2/3}\left(\frac{M}{10^5 \textrm{g}}\right)^{4/3}\Omega_p^{1/3},
\end{equation}
where $\Omega_p$ is the density parameter of PBHs at their formation time. Here the redshift $z$ is defined with respect to the moment of PBH evaporation, so $z=0$ corresponds to 
$t = t_{ev}$, as it is written above eq. \eqref{integral}. It should not be confused with the usual definition of the redshift with respect to the present day. According to estimates of \cite{hi-f-gw} the density parameter of PBH at production depends implicitly on the initial magnitude of the density perturbations during the PBH-dominated era before bing bang nucleosynthesis BBN, $\Delta_i$, and on the effective number of produced particles species, $N_{eff}$. For  reasonable values  $\Delta_i\sim 10^{-4}$ and $N_{eff}\sim 100$, the density parameter of PBHs at production can be in the range $7\cdot 10^{-11}\lesssim\Omega_p\lesssim 1$.

\begin{figure}[htbp]
\begin{subfigure}[b]{0.3\textwidth}
\centering
\includegraphics[scale=0.85]{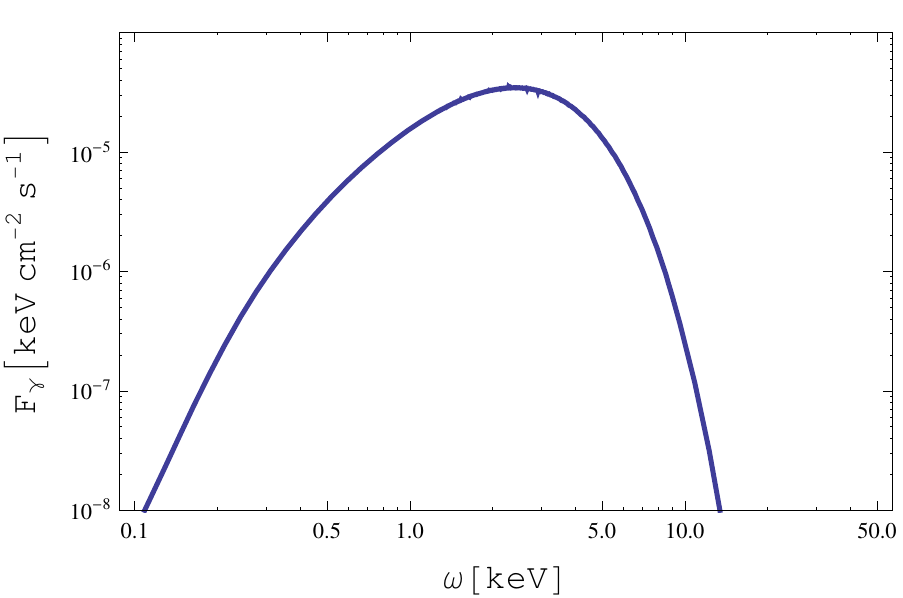}
\caption{}
\label{fig:Fig9}
\end{subfigure}\hspace{3.1cm}
\begin{subfigure}[b]{0.3\textwidth}
\centering
\includegraphics[scale=0.85]{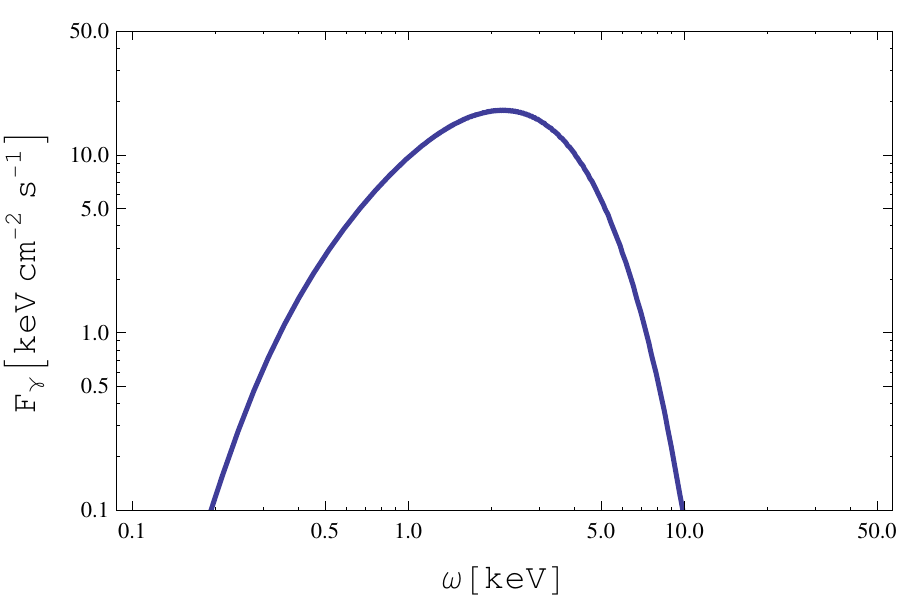}
\caption{}
\label{fig:Fig10}
\end{subfigure}     
\caption{Energy flux, $F_{\gamma}$, of photons produced by the transformations of gravitons into photons as a function of 
the photon energy $\omega$ at the present cosmological epoch, $a=1090$, for the initial value of magnetic field 
$B_i\simeq 3\cdot 10^{-3}$ G (a) and $B_i\simeq 2.37$ G (b).}
\label{fig:Fig6a}
\end{figure}

 In Fig. \ref{fig:Fig6a} the present day energy flux of photons, produced through graviton-to-gamma conversion, $F_\gamma(\omega)=\rho_{\gamma\gamma}(\omega)\cdot F_\textrm{gw}(\omega)$, is presented for 
 PBHs with mass $M_\textrm{BH}\simeq 10^8$~g, $N_{eff}\sim 100$, and $\Omega_p=10^{-3}$. 
 One can  see that these photons can make a
 a substantial contribution to the cosmic x ray background (CXB) in the case of the initial value of magnetic field $B_i\simeq 2.37$ G which corresponds to the
 present day field $B(t_0)\simeq 2\cdot 10^{-6}$ G. 
 The maximum of the spectrum is approximately situated at
 $\omega_{max}\simeq 2.8\, T_\textrm{BH0}$ and corresponds to the photon energy $\omega_{max}\simeq 2.25$ keV 
 and to the flux $F_\gamma\simeq 18$ keV/cm$^2$ s$^{-1}$. 
 In the case of $B_i=3\cdot 10^{-3}$ G the peak of the flux is several orders of magnitude smaller and has the approximate magnitude $F_\gamma\simeq 3\cdot 10^{-5}$ keV/cm$^2$ s$^{-1}$.

\begin{figure}[!h]
\begin{center}
\includegraphics[scale=0.5]{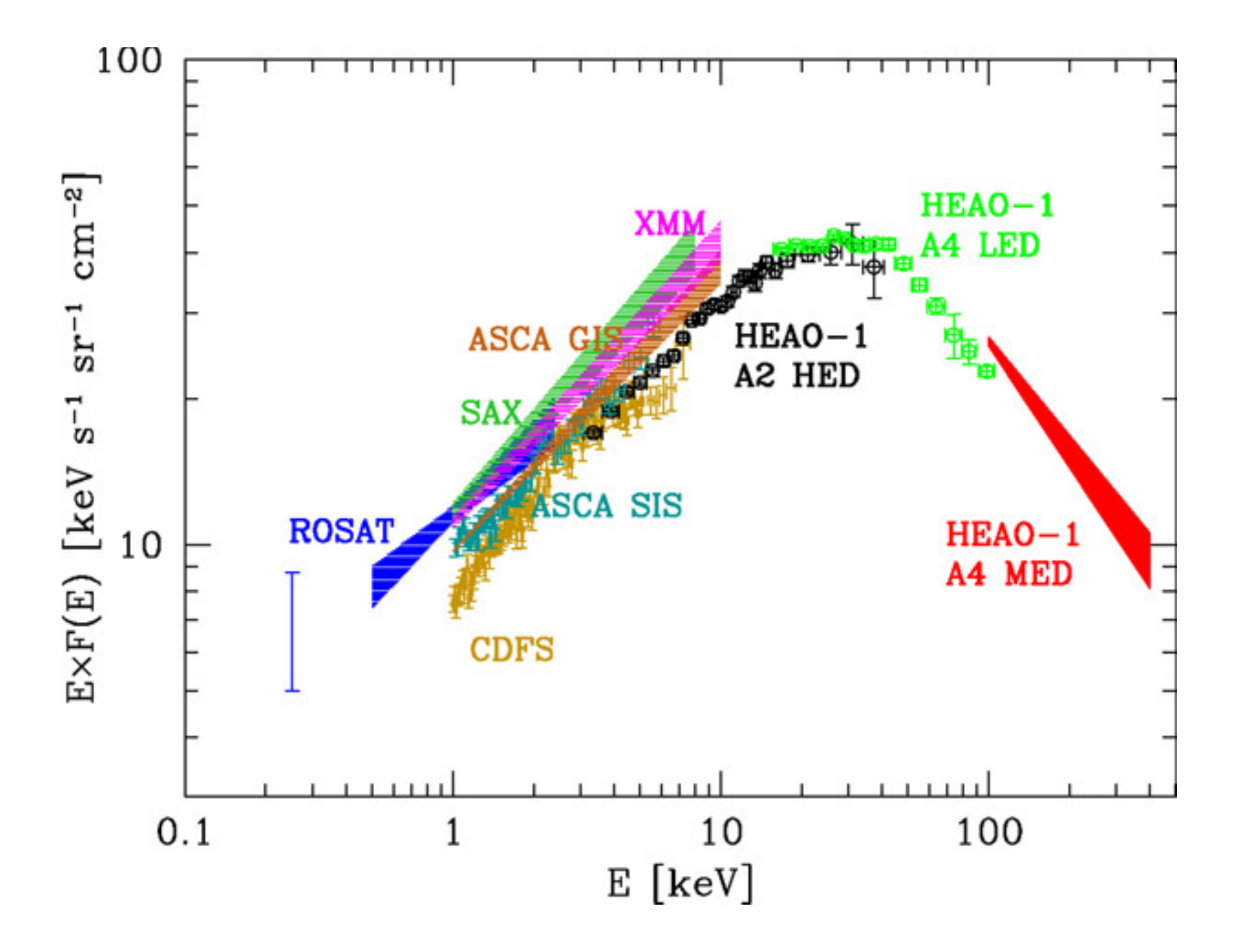}
\caption{Energy flux of the extragalactic X-ray background spectrum from 0.2 to 400 keV as given in \cite{Gilli:2003bm}. Most of the energy flux is concentrated in the energy band of 10-100 keV where the spectrum presents a peak at 30 keV with flux of approximately of 40 keV/cm$^2$ s$^{-1}$.}
\label{fig:Fig7a}
\end{center}
\end{figure}

\section{Discussion and conclusion}\label{sec:7}

We have studied  resonant conversion of gravitons into photons in large scale magnetic fields at the post-recombination epoch. 
The calculated photon production probability is several orders of magnitude higher than that found in the nonresonant case \cite{ad-de-2}.
The result depends on several factors such as the value of the large scale magnetic field $B$, the particle number density in plasma, the position of the resonance, 
and on the coherence breaking effects induced by the Compton scattering on electrons and pair production on atomic electrons and nuclei.

Even in the case of very high graviton energy, which may exceed the electron mass by many orders of magnitude, the nonlinear QED corrections are reduced to the local Euler-Heisenberg effective Lagrangian. The resonance transition is possible for sufficiently high graviton energy which is given by eq. \eqref{resonance-freq}. At the resonance the graviton-photon mixing has the maximum value corresponding to the mixing angle $\theta=\pi/4$. 
This can be seen e.g. in Fig. \ref{fig:Fig3a} where a rapid increase of the transition probability takes place relatively early after recombination. This is an interplay of two effects: an occurrence of the resonance and a rapid decrease of the ionization fraction in the plasma.

The density of the produced photons, apart from the above-mentioned factors,
depends also on the efficiency of the graviton production prior to the hydrogen recombination. Here we consider only graviton production by PBHs evaporated before big bang nucleosynthesis. These black holes are not constrained by any known observational effect \cite{Carr:2009jm} and might emit a substantial background of GWs \cite{hi-f-gw}. In Figs. \ref{fig:Fig4a} and  \ref{fig:Fig5a} the photon production probability is depicted 
as a function of the photon energy at the present time.  It may reach a very high value,
such as $\rho_{\gamma\gamma}\sim 10^{-3}$. 
In Fig. \ref{fig:Fig6a} the spectrum of the produced photons is presented for 
the case of PBHs with the mass  $M_{BH}\sim 10^8$ g.

Depending on the value of the magnetic field at recombination and on the PBH mass, the produced photons could make 
an essential contribution to the extragalactic background light (EBL) and to the CXB. 
If the PBH mass is about 
$M_\textrm{BH}\sim 10^8$ g and $B_i\sim$  a few gauss, the produced  electromagnetic radiation could be the dominant 
component of CXB in the energy band  $0.1-10$ keV see Fig. \ref{fig:Fig6a} and Fig. \ref{fig:Fig7a}. 
For $B_i$ of the order of  a few milli-Gauss and $M_\textrm{BH}\sim 10^8$ g  the contribution to the CXB is smaller but still comparable 
with the flux from some AGN, averaged over the sky. 
For lighter PBHs (for example, for  $M_\textrm{BH}\sim 10^5-10^7$ g) the spectrum if shifted to the lower part of CXB and 
to the ultraviolet 
but  the production probability in that energy range is small in comparison with the resonant case. 
In fact, the peak energy is proportional to the PBH temperature essentially through the combination $\omega_{max}\propto T_\textrm{BH0}\propto (M_\textrm{BH}/N_{eff})^{1/2}$ keV$^{1/2}$ due to the redshift of the graviton energy from the evaporation moment to the present time \cite{hi-f-gw}. So in principle, the heavier is the black hole is, the greater the peak energy is i.e. the energy at which  the flux is maximal. If one assumes that the contribution to $N_{eff}$ comes only from the standard model with $N_{eff}\sim O(10^2)$ and requires that PBHs evaporate before BBN, 
the maximum allowed mass of PBH is about $10^8$ g. The gravitons evaporated by such a PBH would create photons with the
spectrum presented in  Fig. \ref{fig:Fig6a}. On the other hand 
if extra contributions beyond the standard model are taken into account with $N_{eff}\geq 10^3$, the maximum mass of PBH the allowed by BBN can be higher than  $10^8$ g. Since the redshifted PBH temperature is $T_\textrm{BH0}\simeq 0.2\, (M_\textrm{BH}/10^5 \textrm{g})^{1/2}\,N_{eff}^{-1/2}$ keV and knowing that the PBH mass is bounded from above by $M_\textrm{BH}\lesssim 3.88\cdot 10^7\, N_{eff}^{1/3}$ g \cite{hi-f-gw} and that $T_\textrm{BH0}\lesssim 4\, N_{eff}^{-1/3}$ keV, one can conclude that the higher is the effective number of particle species with masses below or comparable to $T_\textrm{BH}$ is, the smaller the PBH temperature should be in order to 
evaporate before BBN. Consequently the number of particle species, is higher, and the redshifted PBH temperature at present $T_\textrm{BH0}$ is lower. So if we assume for example that $N_{eff}\sim 10^5$, the maximum value of the PBH mass from BBN is about $M_\textrm{BH}\sim 10^9$ g 
and the ratio of $M_\textrm{BH}/N_{eff}$ is smaller in comparison with the case of a PBH with a mass of $M_\textrm{BH}\sim 10^8$ g and $N_{eff}\sim 100$. This would imply that for a BH with a mass of $M_\textrm{BH}\sim 10^9$ g the spectrum of produced photons is shifted to the 
lower part of CXB and ultraviolet because its redshifted temperature decreases. This is a consequence of the BBN bound on the maximum PBH mass. 

The model proposed in this article can also be used to constrain several parameters if others are known. Based on the predicted energy flux of this mechanism and using observation of the CXB together with observation in other energy bands where the graviton to photon conversion is efficient, it is possible to constrain either the magnitude of cosmological magnetic field $B$ or the PBH mass and its density parameter at production $\Omega_p$.
\\[3mm]
\noindent
{\bf Acknowledgments.}
We acknowledge the support of the Russian Federation government, Grant No. 11.G34.31.0047.

  \end{document}